\def\vect#1{\bm{\mathrm{#1}}}

\documentclass [twocolumn,superscriptaddress,showpacs]{revtex4}
\usepackage {graphicx}
\usepackage{bm}

\begin{document}

% You may use Title,Subject,Author,Manager,Company,Operator,
% Category,Comment,Hlinkbase document properties here
\title{Holographic analysis of diffraction structure factors}

\begin{abstract}
We combine the theory of inside-source/inside-detector x-ray fluorescence 
holography and Kossel lines/x ray standing waves in kinematic approximation 
to directly obtain  the
phases of the diffraction structure factors. The influence of Kossel lines 
and standing waves on holography is also discussed. 
We obtain partial phase determination from experimental data obtaining 
the sign of the real part of the structure factor for several reciprocal 
lattice vectors of a vanadium crystal. 
\end{abstract}

\date{\today}

\preprint{LBNL}
 \author{S. Marchesini}
%\email{smarchesini@lbl.gov}
\affiliation{Lawrence Berkeley National Laboratory, Berkeley, California, 94720}
\author{N. Mannella}
\author{ C. S. Fadley}
\author{ M. A. Van Hove} 
\affiliation{Lawrence Berkeley National Laboratory, Berkeley, California, 94720}
\affiliation{Department of Physics, University of California at Davis, Davis, California 95616}
\author{J. J. Bucher}
\author{D. K. Shuh}
\affiliation{Lawrence Berkeley National Laboratory, Berkeley, California, 94720}
\author{L. Fabris} 
\author{ M. H. Press}
\author{ M. W. West}
\affiliation{Lawrence Berkeley National Laboratory, Berkeley, California, 94720}
\author{W. C. Stolte}
\affiliation{Department of Chemistry, University of Nevada, Las Vegas, Nevada 89154-4003}
\author{ Z. Hussain}
\affiliation{Lawrence Berkeley National Laboratory, Berkeley, California, 94720}
\maketitle

The phase determination of diffracted beams is the central problem of x-ray 
crystallography. Several methods exist to obtain this information, such as 
direct methods \cite{Sayre:1995}, 3-beam diffraction \cite{Shen:1987}, 
anomalous diffraction, heavy atom or molecular replacement 
\cite{Blundell:1976} and x-ray standing waves or Kossel lines 
\cite{von:1960,Batterman:1964,Cowley:1964}. Despite the 
many advances in these methods, not all problems can be solved. Direct 
methods fail when the unit cell contains a large number of atoms. Anomalous 
diffraction and related methods are among the most successful methods for 
biological crystallography, but they often require chemical modification of 
the molecules. Multiple beam diffraction, x-ray standing waves and Kossel 
lines have usually been applied only to high-quality crystals of relatively 
simple structures, or to the localization of dopants in high quality 
crystals \cite{Golovchenko:1974}. 

 Kossel lines are formed when a 
source of short wavelength radiation ($\sim 1$ \AA) is located on a 
crystallographic site: they result from the Bragg scattering of outgoing 
fluorescent x-rays from various sets of planes in the lattice. In the 
notation of holography, this is an ``inside-source'' experiment. The fine 
structure of these lines has been explained by the dynamical theory of x-ray 
diffraction via the reciprocity theorem used in optics 
\cite{von:1960}. A proper analysis of the KL fine structure allows 
the determination of the phases of reflections 
\cite{Batterman:1964,Cowley:1964,Gog:1995,Vartanyants:2001,Lonsdale:1949}. In parallel to this work on KL, the x-ray standing 
wave (XSW) method has been developed 
\cite{Batterman:1964,Cowley:1964,Golovchenko:1974,Vartanyants:2001}. 
In this case, the source and the detector are interchanged as compared to 
the KL method: the atoms are subject to the changing wave-field in the 
crystal as the incident beam goes through a Bragg reflection, and 
fluorescent radiation proportional to the field at the atom is generated. 
This constitutes the "inside-detector" configuration in holography 
\cite{Gog:1995}. Besides the fine structure of the KL or XSW 
produced when the crystal orientation satisfies a Bragg or Laue condition, 
tails are formed far from the Bragg angle. These coarse features, also 
formed by poorer-quality mosaic crystals provide information on the real and 
imaginary part of the structure factor \cite{Vartanyants:2001,Lonsdale:1949}.

Unlike standard imaging methods, holography offers the possibility of 
extracting both intensity and phase information. X-ray fluorescence 
holography (XFH) is thus a very promising new method for obtaining a direct 
image in real space of the local environments of different atomic species in 
reasonably well-ordered crystals or molecular ensembles. Long-range translational 
order is not required, and indeed one of the most important results obtained by XFH has been the imaging of the average local environment of a quasicrystal
\cite{Marchesini:2000}. Despite this, with the exception of quasicrystals,
 all systems measured so far have been  well-ordered crystals 
\cite{Tegze:1999,Hiort:2000,Tegze:2000,Kopecky:2001}.

In this paper we develop a theoretical method for analyzing the 
inside-source and inside-detector holograms of a periodic object using the 
kinematic approximation. By analyzing the holographic reconstruction in 
reciprocal space, we discuss how to obtain directly the phase of the 
structure factors. We also show how the standard holographic analysis is 
affected by diffraction, and discuss the solution to this problem. 

A hologram is formed whenever an unknown object wave $E_{obj} $ is 
coherently added to a reference wave $E_{ref} $:
\begin{eqnarray}
\nonumber
 I\left( \vect {k} \right) &=& \left| {E_{\text{ref}} + E_{\text{obj}} } \right|^{2} \\ 
\label{eq1}
 &=& I_{\text{ref}} + I_{\text{obj}} + 2\mathrm{Re}\left\{ {E_{\text{ref}}^{\ast } E_{\text{obj}} } 
\right\}. 
\end{eqnarray}
\noindent
where $I_{ref} $ and $I_{obj} $ are the intensities of the reference and 
object waves respectively.

The hologram is extracted by subtracting the reference beam intensity 
$I_{ref} $ and normalizing. The object term $I_{obj} $ is usually assumed to 
be small and therefore neglected in order to permit a holographic analysis 
of the data in XFH. We will see later how this approximation can affect our 
analysis. When an atom located at the origin emits (inside source) or 
detects (inside detector) radiation, the resulting hologram (considering 
only the last term in eq. (\ref{eq1}), i.e. the interference term) can be expressed 
as \cite{Hiort:2000}:
\begin{equation}
\label{eq2}
\chi \left( \vect {k} \right) = - 2\mathrm{Re}\sum\limits_{\vect {r}} {f(\vect 
{k},\vect {r})\,} \,\chi _{\vect {r}} \left( \vect {k} \right),
\end{equation}
\noindent
where $\chi _{\vect {r}} \left( \vect {k} \right) \equiv \, - 
\textstyle{{e^{i\left( {kr - \vect {k} \cdot \vect {r}} \right)}} \over r}$ 
and $f(\vect {k},\vect {r})$ is the scattering factor (including the Thompson 
scattering factor \cite{Len:1997}) of the atom located at $\vect {r}$. The 
summation is extended to all the atomic positions. 

Ideally, when the object term can be neglected and the scattering factors 
are isotropic, the holograms $\chi _{\vect {r}} \left( \vect {k} 
\right)$, $\quad \chi _{{\vect {r}}'} \left( \vect {k} \right)$ generated by two point scatterers located at $\vect {r}$ and ${\vect {r}}'$, 
are orthogonal. \textit{i.e.} $\quad \left\langle {\chi _{\vect {r}}^{\ast } \left( \vect {k} 
\right)\,\chi _{{\vect {r}}'} \left( \vect {k} \right)} \right\rangle _{\vect 
{k}} \simeq$$\alpha \left( \vect {r} \right)\delta \left( {\vect {r} - {\vect 
{r}}'} \right)$. $\alpha \left( \vect {r} \right)$ is a normalization 
function and $\left\langle \right\rangle _{\vect {k}} $ is the average on the 
measured k space. The two holograms are orthogonal because the relative 
phase oscillates when the infinite k-space is spanned. In normal situations, 
due in particular to the limited k-space sampling, $\delta \left( {\vect {r} 
- {\vect {r}}'} \right)$ becomes a point spread function peaked in $\vect {r} 
= {\vect {r}}'$. The typical reconstruction algorithm is based on this 
orthogonality assumption. The holographic reconstruction is the projection 
of $\chi \left( \vect {k} \right)$ onto $\chi _{{\vect {r}}'} \left( \vect {k} 
\right)$:
\begin{equation}
\label{eq3}
U\left( {\vect {r}}' \right) = \left\langle {\chi \left( \vect {k} \right)\chi 
_{{\vect {r}}'}^* \left( \vect {k} \right)} \right\rangle _{\vect {k}} \,.
\end{equation}
What happens if we now perform the reconstruction taking into account the 
long-range periodicity of the system? To answer this question, let us first 
examine the holographic reconstruction in the reciprocal space, by applying 
a Fourier transform to eq. (\ref{eq3}). Since only the kernel $\chi _{{\vect {r}}'} 
\left( \vect {k} \right)$ depends on ${\vect {r}}'$ we can bring the FT inside 
the average and write: 
\begin{eqnarray}
\label{eq4}
 G({\vect {h}}') &=& \left\langle {\chi \left( \vect {k} \right)\int {d^{3}\vect 
{r}\,e^{i{\vect {h}}' \cdot \vect {r}}\,} \chi^* _{\vect {r}} \left( \vect {k} 
\right)} \right\rangle _{\vect {k}} \\ 
\nonumber
 &=& \left\langle {\chi \left( \vect {k} \right)\tilde {\chi}^*_{{\vect {h}}'} 
\left( \vect {k} \right)} \right\rangle _{\vect {k}} \;, \\ 
\label{eq5}
\tilde {\chi }_{\vect {h}} \left( \vect {k} \right) &=& \int {d^{3}\vect 
{r}\textstyle{{ - 1} \over r}e^{i\left\{ {kr - (\vect {k} - \vect {h}) \cdot 
\vect {r}} \right\}}\,} = \textstyle{{ - 1} \over {\left( {\left| {\vect {k} - 
\vect {h}} \right|^{2} - k^{2}} \right)}}\;.
\end{eqnarray}
The formal analogy between (\ref{eq3}) and (\ref{eq4}) suggests we consider 
$\tilde {\chi }_{\vect {h}} \left( \vect {k} \right)$ as the hologram 
generated by a point 
structure factor, located at position $\vect {h}$ of the reciprocal space, 
$i.e.$ a sinusoidal charge density distribution with unitary scattering factor 
amplitude. Similarly to the discussion above, we expect the holograms 
$\tilde {\chi }_{{\vect {h}}'} \left( \vect {k} \right),\;
\tilde {\chi }_{\vect {h}} \left( \vect {k} \right)$ 
generated by two points $\vect {h}$ 
and ${\vect {h}}'$ in the reciprocal lattice to be orthogonal in k-space, 
$i.e. \quad \left\langle {\tilde {\chi }_{{\vect {h}}'}^{\ast } 
\left( \vect {k} \right)\tilde {\chi}^*_{\vect {h}} \left( \vect {k} \right)}
 \right\rangle 
\simeq \beta \left( \vect {h} \right)\delta \left( {\vect {h} - {\vect {h}}'} 
\right)$. In real situations the Dirac delta function is replaced by a 
function peaked in $\vect {h}$.

We now examine the properties of the hologram $\chi \left( \vect {k} \right)$ 
in terms of the structure factors of the reciprocal lattice vectors. 
Following \cite{Adams:1998}, we rewrite (\ref{eq2}) in terms of the electron 
density $\rho \left( \vect {r} \right)$:
\begin{equation}
\label{eq6}
\chi \left( \vect {k} \right) = 2\mathrm{Re}\int {d^{3}\vect {x}\,\rho (\vect 
{r})} \,\,\textstyle{{ - r_{e} } \over r}e^{i(kr - \vect {k} \cdot \vect 
{r})}\;,
\end{equation}
\noindent
where $r_{e} $ is the classical electron radius. For simplicity we have 
approximated the Thompson scattering factor to be a constant. If the system 
is periodic, the charge density distribution can be expressed in a Fourier 
series in terms of the reciprocal lattice vectors $\vect {h}$ and the 
relative structure factors $F_{\vect {h}} $, namely $-r_{e} \rho \left( \vect 
{r} \right) = \sum\nolimits_{\vect {h}} {F_{\vect {h}} } e^{i\vect {h} \cdot 
\vect {r}}$, with $F_{\vect {h}} = \textstyle{{-r_{e} } \over V}\int\limits_V 
{\rho \left( \vect {r} \right)\;e^{i\vect {h} 
\cdot \vect {r}}d} ^{3}\vect {r}$, where $V$ is the unit cell volume. 
The origin, which determines the phase shifts in the structure factors, 
is at the emitting atom in the unit cell.

The hologram (\ref{eq6}) becomes, in analogy with (\ref{eq2}):
\begin{equation}
\label{eq7}
\chi \left( \vect {k} \right) =  2\mathrm{Re}\sum\limits_{\vect {h}} {F_{\vect 
{h}} \tilde {\chi }_{\vect {h}} \left( \vect {k} \right)\;} .
\end{equation}
The divergence of $\tilde {\chi }_{\vect {h}} \left( \vect {k} \right)$ at the 
Bragg condition ($\vect {h}^{2} - 2\vect {h} \cdot \vect {k} = 0)$ in 
(\ref{eq5}) is 
introduced as a consequence of neglecting extinction, mosaicity and the 
finite sample size. These effects can be approximated by writing $k$ as a 
complex number \cite{The:1}, i.e. $k = k_{r} + ik_{i} $, and $\tilde 
{\chi }_{\vect {h}} \left( \vect {k} \right) = \mathrm{Re}\left\{ {\tilde 
{\chi }_{\vect {h}} \left( \vect {k} \right)} \right\} + i\mathrm{Im}\left\{ 
{\tilde {\chi }_{\vect {h}} \left( \vect {k} \right)} \right\}$ then becomes:
\begin{eqnarray}
\nonumber
 \mathrm{Re}\left\{ {\tilde {\chi }_{\vect {h}} \left( \vect {k} \right)} 
\right\} &=& \left| {\tilde {\chi }_{\vect {h}} \left( \vect {k} \right)} 
\right|^{2}\left( { \vect {h}^{2}-2\vect {h} \cdot \vect {k}} \right), \\ 
\nonumber
 \mathrm{Im}\left\{ {\tilde {\chi }_{\vect {h}} \left( \vect {k} \right)} 
\right\} &=& \left| {\tilde {\chi }_{\vect {h}} \left( \vect {k} \right)} 
\right|^{2}\left( {-2k_{i} k_{r} } \right), \\ 
\label{eq8}
 \left| {\tilde {\chi }_{\vect {h}} \left( \vect {k} \right)} \right|^{2} &=& 
\textstyle{1 \over {\left( {\vect {h}^{2} - 2\vect {h} \cdot \vect {k}} 
\right)^{2} + \left( {2k_{i} k_{r} } \right)^{2}}}\;. 
\end{eqnarray}
As the wave-vector $\vect {k}$ changes across the pole of $\tilde {\chi 
}_{\vect {h}} \left( \vect {k} \right)$, there will be a symmetric and an 
antisymmetric contribution dependent in different ways on the real and the 
imaginary part of the structure factor. Writing $F_{\vect {h}} = 
\mathrm{Re}\left\{ {F_{\vect {h}} } \right\} + i\,\mathrm{Im}\left\{ {F_{\vect 
{h}} } \right\}$, the hologram in (\ref{eq7}) becomes:
\begin{equation}
\label{eq9}
\chi =  2\sum\limits_{\vect {h}} {\mathrm{Re}\left\{ {\tilde {\chi }_{\vect 
{h}} } \right\}\mathrm{Re}\left\{ {F_{\vect {h}} } \right\}} - 
\mathrm{Im}\left\{ {\tilde {\chi }_{\vect {h}} } \right\}\mathrm{Im}\left\{ 
{F_{\vect {h}} } \right\} \,.
\end{equation}
Note that, at the Bragg condition, the object term cannot be neglected, it 
contributes with a second order divergence and therefore needs to be added 
to the interference term. The normalized object term can then be written as:
\begin{equation}
\label{eq10}
\frac{I_{obj} }{I_{ref} } = \sum\limits_{\vect {h}} {\left| {F_{\vect {h}} } 
\right|^{2}\left| {\tilde {\chi }_{\vect {h}} } \right|} ^{2} + 
\sum\limits_{\vect {h},{\vect {h}}' \ne \vect {h}} {F_{\vect {h}}^{\ast } 
F_{{\vect {h}}'} \tilde {\chi }_{\vect {h}}^{\ast } \tilde {\chi }_{{\vect 
{h}}'} } \;.
\end{equation}
Except when multiple Bragg conditions are satisfied simultaneously, the 
second term can be neglected. We recall that the object wave intensity is 
always smaller than that of the reference at a pole $\textstyle{{I_{obj} } 
\over {I_{ref} }} \approx \left| {\textstyle{{F_{\vect {h}} } \over {\left( 
{2k_{r} k_{i} } \right)}}} \right|^{2} < 1$, while the interference term 
(\ref{eq9}), at a Bragg condition, is $\textstyle{{\mathrm{Im}F_{\vect {h}} } \over 
{k_{r} k_{i} }}$. The interference term (\ref{eq9}) can still be the dominant one, 
depending on the phase of the structure factor. 

Let us now consider the properties of the reconstructed image $G$, 
which can be 
viewed as the `hologram of the reciprocal lattice', in relation to the 
structure factors. The real and imaginary parts of the reconstruction kernel 
$\tilde {\chi }_{{\vect {h}}'} \left( \vect {k} \right)$ (eq. (\ref{eq8})) are 
respectively antisymmetric and symmetric across the pole as $\left( {{\vect 
{h}}'^{2} - 2{\vect {h}}' \cdot \vect {k}} \right)$ changes from positive to 
negative. The product between an antisymmetric and a symmetric function is 
antisymmetric, and the integral across the pole cancels out:
\begin{eqnarray}
\nonumber
 \left\langle {\mathrm{Re}\left\{ {\tilde {\chi }_{\vect {h}} \left( \vect {k} 
\right)} \right\}\mathrm{Re}\left\{ {\tilde {\chi }^*_{\vect {h}} \left( \vect 
{k} \right)} \right\}} \right\rangle _{k} &=& a\left( \vect {h} \right), \\ 
\nonumber
 \left\langle {\mathrm{Re}\left\{ {\tilde {\chi }_{\vect {h}} \left( \vect {k} 
\right)} \right\}\mathrm{Im}\left\{ {\tilde {\chi}^*_{\vect {h}} \left( \vect 
{k} \right)} \right\}} \right\rangle _{k} &\approx& 0, \\ 
 \left\langle {\mathrm{Im}\left\{ {\tilde {\chi }_{\vect {h}} \left( \vect {k} 
\right)} \right\}\mathrm{Im}\left\{ {\tilde {\chi }^*_{\vect {h}} \left( \vect 
{k} \right)} \right\}} \right\rangle _{k} &=&- b\left( \vect {h} \right); 
\label{eq11}
 \end{eqnarray}
\noindent
where the normalization functions $a\left( \vect {h} \right)$ and $b\left( 
\vect {h} \right)$, are positive functions. The resulting holographic 
reconstruction becomes:
\begin{eqnarray}
\nonumber
 \mathrm{Re}\left\{ {G\left( {{\vect {h}}' = \vect {h}} \right)} \right\} 
&\approx&  {\mathrm{Re}\left\{ {F_{\vect {h}} } 
\right\}a\left( \vect {h} \right)} \,, \\ 
\nonumber
 \mathrm{Im}\left\{ {G\left( {{\vect {h}}' = \vect {h}} \right)} \right\} 
&\approx&  {\left( {2\mathrm{Im}\left\{ {F_{\vect {h}} } 
\right\} - \textstyle{{\left| {F_{\vect {h}} } \right|^{2}} \over {2k_{r} 
k_{i} }}} \right)b\left( \vect {h} \right)} \;. 
 \end{eqnarray}

The real part of the reconstructed hologram, with ${\vect {h}}'$ equal to a 
reciprocal lattice vector $\vect {h}$, is proportional to the real part of 
the structure factor. The characterization of $\mathrm{Re}\left\{ {F_{\vect 
{h}} } \right\}$ reduces the phase problem of the structure factor from the 
possible $(0,2\pi )$ range to only two possible values of the phase, and in 
the centrosymmetric systems this correspond to the complete solution. This 
requires us to calculate the normalization function $a\left( \vect {h} 
\right)$, which is dependent on the measured k-space, polarization factors, 
extinction, sample boundaries, mosaic spread and angular resolution. Such 
calculation goes beyond the scope of this work; however the knowledge of the 
sign of $\mathrm{Re}F_{\vect {h}} $ reduces the range of the possible phases 
by half. The imaginary part of $G$ depends on the real and imaginary part of 
the structure factors. If the object term $I_{obj} $ in (\ref{eq1}) cannot be 
neglected, the Fourier transform of $G,$ the ``standard holographic 
reconstruction'', would have the wrong interpretation. It would be more some 
kind of Patterson function (the Fourier transform of $\left| {F_{\vect {h}} } 
\right|^{2})$ rather than a hologram (Fourier transform of 
$\mathrm{Re}F_{\vect {h}} + i\mathrm{Im}F_{\vect {h}} )$. Only with the full 
knowledge of the normalization functions $a\left( \vect {h} \right)$ and 
$b\left( \vect {h} \right)$ can we separate and extract the values of the 
real and imaginary parts of the structure factors. 
The limitations on this straightforward analysis of the imaginary part 
of the structure factor apply to the standard holographic reconstruction as 
well. It is not surprising that a diffraction measurement interpreted as a hologram in a recent article \cite{Kopecky:2002} provides the correct structure, 
 as the Patterson function peaks match the correct solution.

 Analyzing the reconstruction in the reciprocal space allows 
extracting directly information on the real part of the structure factors, 
simplifying the problem to the calculation of the normalization functions. 
By including the knowledge of the absolute value of the structure factor, we 
can separate the contribution of the object intensity term to the imaginary 
part of the structure factor.
\begin{figure}[htbp]
\centerline{\includegraphics[width=2in]{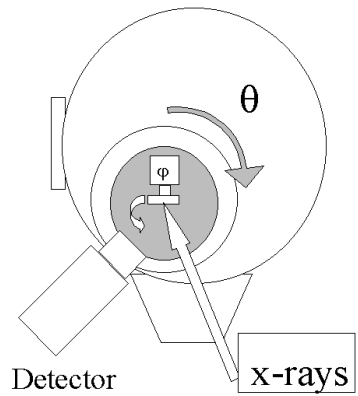}}
\caption{- Experimental setup. Monochromatic x-rays impinge on the sample, mounted on 
a two-axis goniometer that is rotated at high speed, and a solid-state 
detector collects the fluorescence radiation. Only the sample rotates, with 
the angle between the exciting x-rays and the detected fluorescent x-rays 
fixed.}
\label{fig1}
\end{figure}
\begin{figure}[htbp]
\centerline{\includegraphics[width=2.5in]{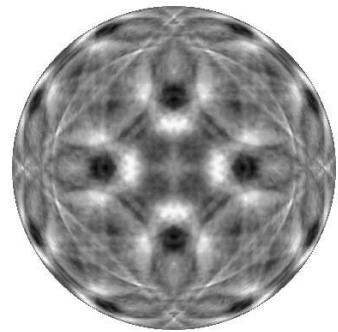}}
\caption{- Measured inside-detector hologram of a vanadium (bcc) single crystal at 
4.9 keV fluorescent energy and 6.3 keV incident energy, after symmetrization 
and rotation of the pattern to be in the (100) orientation.}
\label{fig2}
\end{figure}
\begin{figure}[htbp]
\centerline{\includegraphics[width=3.4in]{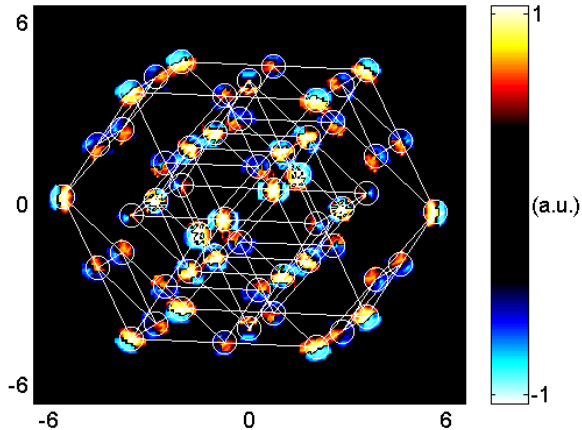}}
\caption{Reconstructed structure factor obtained from the hologram in 
\ref{fig2}; axes scale is in $2\pi / \mbox{{\AA}}$, 
the colorbar is in arbitrary units. Circles indicate the locations of the 
reciprocal vectors, with these being connected by light lines. At each 
circle, there is a positive peak, as expected for this system. }
\label{fig3}
\end{figure}
We performed the experiments on a bending magnet beamline (BL 9.3.1) 
at the Advanced Light Source.
A schematic drawing of the experimental setup is shown in Figure 1. 
Inside the chamber, the sample, a vanadium bcc 
single crystal with (111) orientation, is mounted on a standard two-axis 
 goniometer. The V $K_\alpha $ fluorescence radiation at 4.9 keV emitted 
by the sample was collected by a  4-channel high-speed solid 
state detector with single photon pulse analysis and a maximum of 4 MHz 
count rate \cite{Bucher:1996}. The measurement was performed by rotating the 
sample at high speed in a spiral motion \cite{Marchesini:2001}: the azimuth at 3600 degrees per second, 
and the polar at 2 degrees per second and varying from 0 (perpendicular to 
the surface) to 80 degrees. The detector was placed close to the sample to average the ``inside source'' hologram, resulting in an "inside-detector" measurement in holographic terminology \cite{Gog:1996}. The azimuth stepper motor 
pulses were used to synchronize the data acquisition allowing us to collect 
a full pattern of 3.2 10$^{5}$ pixels in about 40 seconds.
The measurement was repeated until the statistical noise and incident beam 
fluctuations have been reduced to a reasonable level; typically, several 
hundred separate patterns were thus summed in a final dataset.

The measured hologram at 6.3 keV incident energy is shown in Figure 2 and 
the real part of the reconstructed `reciprocal' hologram $\mathrm{Re}\left\{ 
{G\left( {\vect {h}}' \right)} \right\}$ as derived from Eqs. 4 and 5 is 
shown in Figure 3. The network of white lines connects the known positions 
of the reciprocal lattice positions in the V lattice. The image in 
reciprocal space shows its most intense yellow spots at the reciprocal 
lattice positions, which in turn correspond to the positive values of the 
real part of the structure factors. One can see that the height of these 
peaks is approximately constant. This is because the functions $a\left( \vect 
{h} \right)$ and $b\left( \vect {h} \right)$ are almost constant when $\left| 
\vect {h} \right|$ is not close to the wavenumber $k$, this can be true if 
the hologram is measured at a single energy in every direction, as was our 
case. The vanadium crystal is a special case in which the structure factors 
are real and positive. However the reconstructed image from a simulated 
Kossel line pattern of a PbSe single crystal which was chosen since it 
exhibit both positive and negative signs in the structure factors, correctly 
shows these different signs. 

We have presented a method for the direct phase determination of the 
structure factors. This result has been obtained by analyzing the 
holographic reconstruction in reciprocal space and by combining the theory 
of inside-source/inside-detector holography and Kossel lines/x-ray standing 
waves. This method can be applied to any crystal possessing an atom which 
can be excited to emit radiation. We have shown how holograms and standard 
holographic reconstruction can be distorted in periodic objects by x-ray 
diffraction, and discussed the possible solution to this problem. By 
separating the real and imaginary parts of the reconstructed image, and by 
calculating the normalization function, we obtain the real part of the 
structure factor. By including the knowledge of the absolute value of the 
structure factor, one can separate the contribution of the object intensity 
term to imaginary part of the structure factor. 

We have demonstrated this method experimentally on a simple test case, by 
measuring the full inside-detector XSW pattern and obtaining the 
real part of the structure factors for a vanadium crystal. In order to obtain 
the full phase determination, the calculation of the normalization functions 
$a\left( \vect {h} \right)$ and $b\left( \vect {h} \right)$. 
However even the 
qualitative direct image obtained provides the sign of the real part of the 
structure factor, which for centro-symmetric systems correspond to the 
complete solution. The information obtained by this technique can be used 
as input in a standard KL/XSW fitting analysis to obtain the full phase determination.

We would like to acknowledge M. R. Howells and J. C. H. Spence for useful 
discussions and L. Zhao for help during the experiment. 
This work was supported by the U.S. Department of Energy under Contract No.
DE-AC03-76SF00098.


\begin{thebibliography}{17}
\bibitem{The:1} The finite crystal size and the mosaicity produce a broadening of the Fourier coefficients $F_{\vect {h}} $ in the reciprocal space, and the Fourier series ${\rho }'\left( \vect {x} \right) = \sum\nolimits_{\vect {h}} {F_{\vect {h}} } e^{i\vect {h} \cdot \vect {x}}$ needs to be replaced by a Fourier integral.
\end{thebibliography}
\end{document}